\documentclass[ %
12pt,
a4paper,
ngerman,
english,
]{scrartcl} %
\usepackage{babel}              %German language support
\usepackage[utf8]{inputenc}   %German input encoding 
\usepackage[T1]{fontenc}        %Line separation for words with Umlaut
\usepackage[scaled]{helvet}     %Helvetica scaled like Times New Roman
\setcapindent{0em}

\usepackage{amsmath}
\usepackage{amsfonts}
\usepackage{amssymb}
\usepackage{amsthm}    % Definitions, Lemmata, Theorems 
\usepackage{graphicx}   % for figures
\usepackage{caption}
\usepackage{subcaption}
\usepackage[automark]{scrpage2} % for headers
\usepackage{geometry} %for larger headers to avoid overfull hboxes
\usepackage[round,comma,sort]{natbib} %for the round parenthesis around the year
\usepackage{color}
\usepackage{epstopdf}           %To put eps inside the text
\usepackage{fancybox}           %shadowbox, doublebox, ovalbox
\usepackage{tabu}
\usepackage{hhline}
\usepackage[draft]{hyperref}
\hypersetup{final}
\usepackage{cleveref}
\usepackage{makeidx}            %Making indices
\usepackage{enumerate}          %For roman indices in enumerate
\usepackage{float}
\usepackage{doi}
\makeindex

\newif\ifPics %
\Picstrue %\Picstrue \Picsfalse

\begin{document}

% Title
% ------------------------------------------------------------------
\title{Offset errors in probabilistic inversion of small-loop frequency-domain electromagnetic data: a synthetic study on their influence on magnetic susceptibility estimation} %
% Authors
% ------------------------------------------------------------------
\newcommand*\samethanks[1][\value{footnote}]{\footnotemark[#1]}
\author{Christin Bobe%
  \thanks{ %
    Corresponding Author: christin.bobe@ugent.be, %
    Department of Environment, %
    Ghent University, %
    Gent, %
    Belgium %
  } %
  \qquad  Ellen Van De Vijver%
  \\ Department of Environment, %
    Ghent University, %
    \\Gent, %
    Belgium %%
} %
\date{} %
\maketitle

% Preprint statement
% ------------------------------------------------------------------
Accepted for publication in International Workshop on Gravity, Electrical 
and Magnetic Methods and Their Applications (GEM) 2019 Xi'an. 
Copyright (2019) Society of Exploration Geophysicists (SEG) and Chinese Geophysical 
Society (CGS). 
Further reproduction or electronic distribution is not permitted.

\newpage

% Abstract
% ------------------------------------------------------------------
\begin{abstract}
Small-loop frequency-domain electromagnetic (FDEM) devices measure a 
secondary magnetic field caused by the application of a stronger primary 
magnetic field. %
Both the in-phase and quadrature component of the secondary field commonly 
suffer from systematic measurement errors, which would result in a non-zero 
response in free space. %
The in-phase response is typically strongly correlated to subsurface 
magnetic susceptibility. %
Considering common applications on weakly to moderately susceptible 
grounds, the in-phase component of the secondary field is usually weaker 
than the quadrature component, making it relatively more prone to 
systematic errors. %
Incorporating coil-specific offset parameters in a probabilistic 
inversion framework, we show how systematic errors in FDEM measurements 
can be estimated jointly with electrical conductivity and magnetic 
susceptibility. %
Including FDEM measurements from more than one height, the offset estimate 
becomes closer to the true offset, allowing an improved inversion result 
for the subsurface magnetic susceptibility. %
\end{abstract}

\section{Introduction}
\label{sec:introduction}

% General Introduction
% ------------------------------------------------------------------

Electrical conductivity (EC) and magnetic susceptibility (MS) determine 
frequen-cy-domain electromagnetic measurements (FDEM). 
Using small-loop sensors, secondary field responses are usually small 
compared to the present primary field. 
Small measurement quantities are particularly sensitive to systematic 
calibration errors. 
These errors imply that a hypothetic free-space measurement would yield 
non-zero results \cite[]{sasaki2008resistivity}. 
Therefore, one way of calibration is to quantify the offset error by 
performing a measurement far from any conductive matter. 
Unfortunately, this is often practically infeasible. 
Alternatively, the offset is often corrected by repeated measurements on 
multiple elevations above ground, including the offset as an additional 
parameter in an inversion of such data, under the assumption that the 
offset is constant. 
This procedure has been proven to result in more reliable estimates for 
subsurface EC (e.g., \cite{sasaki2008resistivity} and 
\cite{tan2018simultaneous}).
 
% Agenda
% ------------------------------------------------------------------

However, the aforementioned studies focus on EC and do not include 
inversion for subsurface MS. 
The FDEM in-phase response usually shows a strong correlation to subsurface 
MS, but, is often smaller than its quadrature counterpart and therefore even 
more prone to offset errors. 
For that reason, the in-phase response is often not used in inversion for 
EC. 
We show, using a synthetic example, how multi-elevation measurements from 
multi-coil measurement setups can be used to derive reliable MS inversion 
results and an estimation for offset errors. 
The synthetic subsurface model consists of three horizontal layers, 
including variation in EC as well as in MS. 
FDEM responses for this subsurface are simulated using a one-dimensional 
solution to Maxwell’s equations \cite[]{ward1988electromagnetic}. 
Applying the Kalman ensemble generator \cite[]{nowak2009best} for inversion of the 
simulated FDEM responses, we perform a probabilistic inversion, approximating 
probability distributions by an ensemble of subsurface realizations.

Similar to former studies (e.g., \cite{sasaki2008resistivity} and 
\cite{tan2018simultaneous}), we add constant offset errors for each receiver 
coil as parameters in the inversion framework. 
The probabilistic framework allows to jointly estimate EC, MS and the offset 
errors. 
Unlike to the method proposed by \cite{sasaki2008resistivity}, we do not introduce 
further regularization to an iterative inversion process. 
Instead, the offset is estimated through a search of the prior model parameter 
space, resulting in an offset estimate including uncertainty. 
For our synthetic study, we will give offset priors where only in one case 
the true offset matches the prior mean. 

The advantage of multiple elevations for a FDEM measurement inversion is 
demonstrated by comparing the result to an inversion in which only 
measurements from one height are considered. 
Finally, both results are compared to the inversion result for a data set 
without offset errors.

\section{Forward simulation}
\label{sec:fwd}

The synthetic FDEM data are simulated using a one-dimensional forward 
model, allowing for vertical variation of EC and MS (\cite{ward1988electromagnetic}, 
and \cite{hanssens2019fwdemi1D}). 
In small-loop systems, a primary field is computed by assuming an alternating 
current in a transmitter coil. 
A secondary field is simulated based on induction currents in conductive 
model layers. 
The simulated measurement quantity is the field strength as registered by 
the induced voltage in a receiver coil at a defined model position with 
respect to the transmitter coil. 
In the following, the secondary field is expressed in parts-per-million 
(ppm) of the primary field. 
The offset error is included in the forward model by shifting the responses 
by a constant value.

\section{Inversion method}
\label{sec:Inv}

The previously described forward model, in the following called $g$, also serves 
as the kernel for the probabilistic Kalman ensemble generator inversion 
(KEG; \cite{nowak2009best}). %
A full publication on this FDEM inversion procedure is currently under review 
with moderate revisions. %
In this probabilistic framework, we seek a posterior probability distribution 
$\rho(\mathbf{m}|\mathbf{d})$  for the inversion parameters $\mathbf{m}$ given 
FDEM measurements $\mathbf{d}$. % 
According to EC and MS for each discretized layer and a coil specific offset, 
the forward responses are computed:

\begin{equation}
\mathbf{d}_{sim}=g(\mathbf{m}).
\label{eq:fwdsim}
\end{equation}

Inserting the true physical parameters $\mathbf{m}^{true}$ in $g$, we derive 
the synthetic measurements $\mathbf{d}_{obs}=g(\mathbf{m}^{true})$. 
Using Bayes theorem, one derives the posterior probability solving

\begin{equation}
\rho(\mathbf{m}|\mathbf{d})=\kappa \cdot \rho_M(\mathbf{m}) 
\rho_D (\mathbf{d}|\mathbf{m}),
\label{eq:Bayes}
\end{equation}

with a normalization constant $\kappa$ \cite[]{allmaras2013estimating}.  
The likelihood $\rho_D (\mathbf{d}|\mathbf{m})$, for measurements 
$\mathbf{d}$ given a set of parameters $\mathbf{m}$, is computed by comparing 
$\mathbf{d}_{sim}$ to $\mathbf{d}_{obs}$ (assuming a noise-free $\mathbf{d}_{obs}$ 
for the synthetic data). 
The prior information on the Gaussian parameters is given by $\rho_M(\mathbf{m})$ 
(see below). 

As mentioned above, for solving equation \ref{eq:Bayes}, we apply the KEG as 
introduced by \cite{nowak2009best}, a stationary implementation of the widely used 
Ensemble Kalman filter (EnKF; \cite{evensen2003ensemble}). 
The advantage of using a probabilistic approach to inversion lies in the 
simplicity of including the offset error as estimated parameter, and in 
the availability of an uncertainty estimate. 
The EnKF is an ensemble approximation to the original Kalman filter 
\cite[]{kalman1960new}. 
In EnKF, Gaussian probability density functions (PDFs) for model parameters 
are characterized by random samples drawn from these PDFs. 
These $n_{ens}$ random samples are single model realizations which are combined 
in an ensemble matrix $\mathbf{A} \in \mathbb{R}^{(m \times n_{ens})}$, for which $g$ is computed and stored in 
the response matrix $\mathbf{G}\in \mathbb{R}^{(m \times n_{ens})}$. 
Incorporating random noise as standard deviation (STD) to the FDEM measurements 
for all $n_{coils}$  coils, also here $n_{ens}$ samples are drawn to derive the 
data matrix $\mathbf{D}\in \mathbb{R}^{(n_{coils} \times n_{ens})}$. 
The inversion update is derived from the matrix equation \cite[]{evensen2003ensemble}

\begin{equation}
\mathbf{A}^{Update}=\mathbf{A}\mathbf{A'G'^T}(\mathbf{G'G'^T} + 
\mathbf{D'D'^T})^{-1} (\mathbf{D} - \mathbf{G}),
\label{eq:KalmanUpdate}
\end{equation}

where the primed matrices mean that the mean value of each column is subtracted 
from this column.

\paragraph{Prior information} 
The number of inversion parameters is determined by (1) the 
number of discrete subsurface layers, to which we assign EC and MS values, and 
(2) the number of coils, to each of which we assign offsets for both the in-phase 
and the quadrature-phase measurements. 
To all parameters we attribute prior information through definition of their 
prior PDFs. 
To enforce positive values of EC and MS, we use log-normal distributions that 
are transformed to normal distributions before the update step.

To exclude spurious detail and limit the influence of the prior model, we restrict 
the prior to a model with uniform mean and STD for all discrete layers. 
The uniform prior is derived from Gauss sampling, i.e. mean and STD for EC and 
MS are derived considering 10 cm intervals of the synthetic true. 

When no geological layer boundary is present, electric and magnetic properties 
of adjoining layers will be similar. 
A simple covariance matrix $\Sigma$ is introduced in which only the diagonal and 
the first off-diagonals are filled with non-zero values. 
On the diagonal, we write the variance of the expectations $\mu$, on the first 
off-diagonals we include a value smaller than variance as correlation between 
the layers (half of the variance in our case). 
Finally, the prior model is described as a multivariate Gaussian 
$N(\vec{\mu},\Sigma)$.

The prior offsets shall be updated independently for each coil and each 
in-phase and quadrature component. 
Hence, no correlation is introduced in between offset parameters. 
Moreover, no logarithmic barrier is introduced here since the offset is 
assumed to take positive as well as negative values (often dependent on the 
specific coil geometry).

\paragraph{Inversion result} 
The Gaussian posterior is derived by computing the mean and the STD of the 
ensemble realizations for all parameters in $\mathbf{A}^{Update}$. 
We consider the mean as the best fit to the measurement data, while the 
standard deviation is considered as its corresponding uncertainty. 
The Gaussianity of the posterior is enforced by the KEG method, such that 
the derived uncertainty information can only be considered unbiased when the 
forward solutions around $\mathbf{m}^{prior}$ do not deviate too strongly 
from linearity.

\section{Synthetic example}
\label{sec:syn}

\begin{figure}
\centering
\includegraphics[scale=0.35]{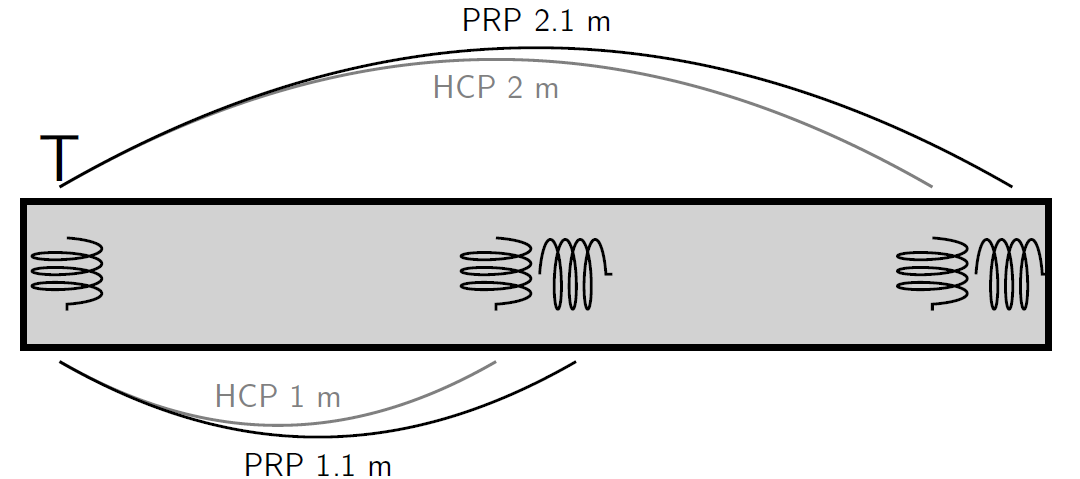}
\caption{Simulated measurement setup: transmitter T with four receiver coils, 
two in horizontal co-planar (HCP) configuration, two in perpendicular 
(PRP) configuration.}
\label{fig:dualem}
\end{figure}

\begin{figure}[ht]
\centering
\includegraphics[scale=0.5]{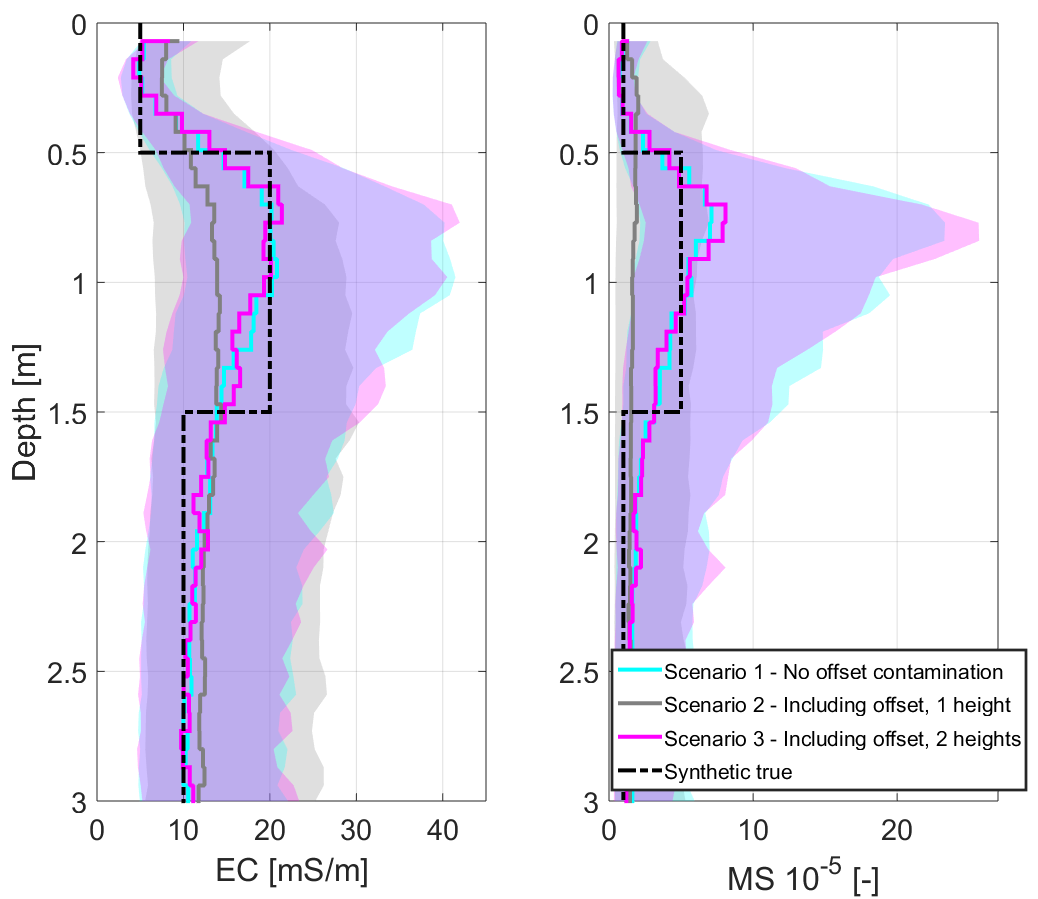}
\caption{Comparison of the best fit models (solid lines) for the three scenarios 
with the synthetic true model for EC ad MS. 
Two posterior standard deviations around the best fit are shown as transparent 
fills in corresponding colors. 
Note their asymmetry due to the assumed lognormal behaviour for EC and MS 
parameters.}
\label{fig:finalcompare}
\end{figure}

The synthetic example consists of a subsurface including three horizontal 
layers, varying in EC as well as in MS (synthetic true shown in Figure \ref{fig:finalcompare}). 
The simulated FDEM measurement setup consists of one transmitter coil and 
four receiver coils (Fig. \ref{fig:dualem}). 
Two receiver coils are in a horizontal co-planar (HCP) configuration, the 
two other receivers in a perpendicular (PRP) configuration. 
Measurements are simulated at two height above ground, at 0.2 m and 1 m 
(Table \ref{tab:measpriorval}). 

Using the outlined KEG inversion, we compare three scenarios. 
Scenario 1 uses the inversion as described above, but without including 
offset parameters, i.e. using data simulated at one height (0.2 m) that are 
not contaminated with an offset shift. 
This is considered the benchmark for the other two scenarios. 
For scenarios 2 and 3, data was shifted with an offset as listed in Table 
\ref{tab:measpriorval}. 
For scenario 2, we only consider measurement data from 0.2 m (Table 
\ref{tab:measpriorval}) height. 
Scenario 3 uses additional simulated measurements at 1 m height. 
Thus, twice as many measurements are used in deriving the inverse model.

Another plausible scenario is not considered in this manuscript: using 
measurements contaminated by offset shifts yet not include the offset in 
the inversion parameters. 
Such a scenario leads to unphysical updates, since the measurement data are 
inherently inconsistent. 
The inconsistency is caused by the fact that the offset remains constant, 
while the measurement response gets weaker as the instrument is lifted. 

\begin{table}[ht]
\centering
\caption{Quadrature- and in-phase responses for FDEM measurements at 0.2 m 
and 1.0 m height. Measurements from both heights are shifted by the "True 
Offset". 
The mean and STD of the Gaussian prior models for the offset estimation are 
listed in the fourth and fifth column.}

\begin{tabular}{l*{6}{c}r}
 & \multicolumn{5}{c}{ \textbf{Quadrature-phase [ppm]}}  \\
 & & & & \multicolumn{2}{c}{\textbf{Prior offset}} \\
 & \textbf{0.2 m} & \textbf{1.0 m} & \textbf{True Offset} & \textbf{mean} 
 & \textbf{STD} \\
  \hline
 HCP 1 m & 183.61 & 85.60 & 13.00 & 18.00 & 3.00 \\
 PRP 1.1 m & 129.18 & 27.96 & -100.00 & -180.00 & 30.00 \\
 HCP 2 m & 832.36 & 550.23 & 24.00 & 22.00 & 5.00 \\
 PRP 2.1 m & 724.37 & 272.41 & -19.00 & -22.00 & 5.00 \\
 \hline
 & \multicolumn{5}{c}{ \textbf{In-phase [ppm]}}  \\
 \hline
 HCP 1 m & 2.99 & 1.93 & 19.00 & 12.00 & 3.00 \\
 PRP 1.1 m & -10.13 & -1.23 & -17.00 & -12.00 & 3.00 \\
 HCP 2 m & 36.82 & 23.63 & 20.00 & 20.00 & 5.00 \\
 PRP 2.1 m & -12.00 & -5.03 & -21.00 & -20.00 & 5.00 \\
 \hline
\end{tabular}
\label{tab:measpriorval}
\end{table}

For EC, MS and the offset, identical prior models are chosen for all 
investigated scenarios – except for the offset prior in scenario 1, since 
there is no offset in this scenario. 
Inverse model layers have a thickness of 0.07 m. 
The STD for the FDEM data matrix are set to a negligibly small value of 
0.05 ppm, expressing that the simulated measurements are considered to be 
noise free. 
We sample the prior PDFs creating an ensemble of 10,000 model realizations.

Except for the HCP 2m coil in-phase response, the chosen prior mean values 
are not identical to the true offset. 
For the quadrature component of the PRP 1.1 m coil, a large offset is given.
It is also noted that the true offset is not within one STD of the prior 
model.
 
The results for the offset estimation are listed in Table \ref{tab:OffsetEstval}. 
The corresponding inversion results for both EC and MS for all scenarios 
are shown in Figure \ref{fig:finalcompare}. 
The best fit offset error estimation for scenario 3 is closer to the true 
offset than for scenario 2, except for the quadrature phase HCP 2 m coil where 
the results are similar. 
It is remarked that the offset uncertainty might be larger when less receiver 
coils are used or when the subsurface is more heterogeneous, also using 
measurements at two (or more) heights. 
In these situations, more elevations per inversion location should be considered 
until the desired precision for the posterior offset standard deviation is 
achieved.

As illustrated in Figure \ref{fig:finalcompare}, the more precise offset 
estimation in scenario 3 also greatly improves the behavior of the best fit 
compared with the best fit for scenario 2. 
The scenario 2 best fit significantly underestimates and smooths the synthetic 
true model geometries. 
This is particularly evident for the MS inverse model, where there is almost 
no contrast resolved between the different MS layers. 
The scenario 3 best fit clearly resembles the best fit for the benchmark 
scenario 1. 
Likewise, derived posterior uncertainties look mostly alike.

\begin{table}[ht]
\centering
\caption{Comparing the true offset for all coils to the modeled best fit 
offset and the corresponding STD (indicated by plusminus term) for the data 
simluated at one height (Scenario 2) and simulated at two heights 
(Scenario 3).}

\begin{tabular}{l*{4}{c}r}
 & \multicolumn{3}{c}{ \textbf{Quadrature-phase [ppm]}}  \\
 & \textbf{True} & \textbf{Scenario 2} & \textbf{Scenario 3}\\
 \hline
 HCP 1 m & 13.00 & 10.21 $\pm$ 1.38 & 13.23 $\pm$ 0.72  \\
 PRP 1.1 m & -100.00 & -115.22 $\pm$ 5.41 & -100.04 $\pm$ 0.07 \\
 HCP 2 m & 24.00 & 26.05 $\pm$ 4.90 & 26.10 $\pm$ 5.70 \\
 PRP 2.1 m & -19.00 & 19.91 $\pm$ 4.71 & -19.32 $\pm$ 0.39 \\
 \hline
 & \multicolumn{3}{c}{ \textbf{In-phase [ppm]}}  \\
 \hline
 HCP 1 m & 19.00 & 14.20 $\pm$ 1.89 & 19.21 $\pm$ 0.34 \\
 PRP 1.1 m & -17.00 & -18.32 $\pm$ 1.66 & -16.95 $\pm$ 0.07 \\
 HCP 2 m & 20.00 & 12.31 $\pm$ 3.71 & 21.47 $\pm$ 2.56 \\
 PRP 2.1 m & -21.00 & -29.87 $\pm$ 1.66 & -20.79 $\pm$ 0.27 \\
 \hline
\end{tabular}
\label{tab:OffsetEstval}
\end{table}

A quantitative comparison of scenario 1, 2 and 3 is somewhat flawed since 
the benchmark scenario 1 and scenario 2 use only four independent FDEM 
measurements, while scenario 3 uses eight. 
Additionally, a less uncertain prior model for the offsets or for subsurface 
EC and MS, might as well improve the posterior inverse model of single-height 
FDEM measurement data.

Although, the in-phase HCP 2 m and PRP 2.1 m measurements got identical 
absolute prior offset shifts, their posterior models are different. 
This might be caused by a fortunate measurement sensitivity with depth for the 
PRP 2.1 m coil in relation to the layer boundaries. 
A detailed discussion of this phenomenon is beyond the scope of the presented 
work and might be discussed later.  

To demonstrate the effect of a rather far-off prior guess, the quadrature-phase 
PRP 1.1 m coil offset is assigned a prior model with mean -180.00 ppm and a STD 
of 30.00 ppm while the true offset is -100.00 ppm. 
The offset best fit with one standard deviation is computed to -115.22 $\pm$5.41 
ppm for scenario 2 and to -100.04 $\pm$0.07 ppm for scenario 3. 
Particularly in the latter, the posterior uncertainty is remarkably small. 
We therefore conclude that offsets estimations are expected to be relatively 
robust towards wrong prior guesses.

\section{Conclusion}
\label{sec:conclusion}

Offset shifts have a relatively stronger effect on posterior MS estimates 
than on EC estimates, when in-phase offset effects are in the same order of 
magnitude as the quadrature-phase offsets. 
Probabilistic inversion of FDEM measurements on two heights including 
coil-specific offset errors as extra inversion parameters allows for a 
reliable offset error estimation. 
Additionally, the posterior estimate for EC and MS is improved. 
In the probabilistic inversion framework, the uncertainty information for 
the best fit can be used as a relative measure for reliability of the offset 
estimates when further, independent FDEM measurements from a coil on different 
heights are added.

% Acknowledgments
% ------------------------------------------------------------------
\section{Acknowledgments}

This project has received funding from the European Union’s EU Framework 
Programme for Research and Innovation Horizon 2020 under Grant Agreement 
No 721185.

% Bibliography
% ------------------------------------------------------------------
%
%\bibliography{ArticleOffset_Preprint}
%\bibliographystyle{apalike}

\end{document}